\documentclass[10pt]{article}
\usepackage[T1]{fontenc}
\usepackage[margin=1.4in]{geometry}
\usepackage{mathptmx}
\usepackage{microtype}
\usepackage{amsmath,amssymb}
\usepackage{amsthm}
\theoremstyle{definition}
\newtheorem{principle}{Principle}
\usepackage{graphicx}
\usepackage{float}
\usepackage{booktabs}
\usepackage{array,tabularx}
\usepackage[numbers,sort&compress]{natbib}
\usepackage{xcolor}
\usepackage[colorlinks=true,linkcolor=blue!55!black,citecolor=blue!55!black,urlcolor=blue!55!black]{hyperref}
\usepackage{caption}
\newcolumntype{Y}{>{\raggedright\arraybackslash}X}

\title{\bfseries A Unifying Perspective on Audio Generative Modeling:\\[2pt] Latent Representations and Modeling Strategies}
\author{Dongchao Yang\thanks{This is a preliminary version. Comments and corrections are welcome.}\\
The Chinese University of Hong Kong\\
\texttt{dcyang@se.cuhk.edu.hk}}
\date{}

\newcommand{\fig}[3]{%
  \begin{figure}[H]\centering
  \includegraphics[width=#2\linewidth]{figs/#1}
  \caption{#3}\label{fig:#1}\end{figure}}

\begin{document}
\maketitle

\begin{abstract}
Every audio generative system makes two coupled decisions: what representation to generate, and how to model its distribution. This paper organizes audio generative modeling around this coupling. For representation design, we compare discrete, continuous, and hybrid latents through four objectives: representation burden, distortion, empirical modelability, and streaming compatibility. For distribution modeling, rather than treating a latent's difficulty as an intrinsic scalar, we use two diagnostic dimensions: dependency horizon, how far useful context extends, and conditional ambiguity, how much uncertainty remains after conditioning. These dimensions refine the common semantic-versus-acoustic intuition: variables with a long dependency horizon should receive global modeling capacity. Conditionally ambiguous detail may be delegated to a local or iterative generator. Applied to representative systems, this view shows that RVQ's residual order gives ordered capacity but not ordered semantics, that AudioLM's semantic-versus-acoustic cascade is one explicit placement of this boundary rather than a universal template, and that autoregression, iterative refinement, and hybrid designs differ chiefly in how they trade dependency horizon against critical-path generation cost. The distinction between discrete and continuous latents describes the output interface; dependency horizon, conditional ambiguity, and streaming determine how that interface should be modeled. Rather than cataloguing individual systems, we provide an evaluation and design framework for comparing representation--model pairs.
\end{abstract}

\section{Introduction}
\label{sec:intro}

Every audio generative system makes two coupled choices: what latent representation to generate, and how to model its distribution. Although these choices are often implemented in separate modules, with an encoder or codec followed by a generative model, they cannot be evaluated independently. The latent determines the sequence length, the information carried at each step, and the dependencies exposed to the generator. The modeling strategy determines how those dependencies are factorized and how the remaining uncertainty is represented. Together, they determine generation quality, computational cost, and whether the system can operate in real time. We aim to provide a structured perspective on audio generative modeling, not an exhaustive catalogue or a theorem about one optimal architecture.

SoundStream~\citep{soundstream} and EnCodec~\citep{encodec} illustrate one form of this coupling. They keep the temporal frame rate low by representing each frame with multiple codebook indices through residual vector quantization (RVQ). The resulting latent has two dependency axes, time and codebook depth, and is no longer a standard one-token-per-step LM target. Flattening both axes preserves rich conditioning but increases serial generation cost. Predicting codebooks in parallel is faster but removes part of their conditional structure. VALL-E~\citep{valle}, MusicGen~\citep{musicgen}, UniAudio~\citep{uniaudio}, and Moshi~\citep{moshi} choose different compromises between these two costs. The codec has therefore not only compressed the waveform; it has also determined the factorization problem faced by the downstream model. AudioLM~\citep{audiolm} illustrates a second form of the same coupling: it shows where a representation draws the boundary between global content and local, ambiguous detail. Semantic tokens make long-range linguistic structure directly predictable but do not preserve enough information for waveform reconstruction. We interpret AudioLM as an explicit allocation: it predicts a semantic trajectory first and generates acoustic tokens conditionally, drawing this boundary between two separate token vocabularies. DiTAR~\citep{ditar} and CALM~\citep{calm} draw it inside a single continuous latent instead. A causal outer process tracks history. Diffusion or a consistency head resolves the remaining acoustic ambiguity within each step or block.

RVQ's codebook depth and a continuous latent's per-token dimensionality buy the same representation burden in two different ways. AudioLM's semantic tokenizer and DiTAR's and CALM's causal-outer, diffusion-inner split draw the same global-versus-local boundary in two different ways. Discreteness and continuity decide neither outcome. What varies across these systems is how dependency structure and conditional ambiguity are allocated, not the representation's category.

These examples motivate the two questions that organize the paper. \textbf{What should an audio model generate?} We compare discrete, continuous, and hybrid latents through four requirements: rate, distortion, modelability, and streaming. \textbf{How should the distribution of that latent be modeled?} We examine how far its dependencies extend, how much ambiguity remains after conditioning, and whether future context is available. These properties determine when long-context autoregression, iterative refinement, codebook-depth modeling, or a hybrid factorization is appropriate.

The analysis proceeds in three steps. We first describe a four-objective representation design space. We then refine the intuition that semantics is low-entropy and acoustics is high-entropy, using two diagnostic questions: which variables benefit from longer context, and which retain multiple acceptable realizations after conditioning? Finally, we analyze how different modeling schemes assign these variables to long-context, local, or iterative models. Appendix~\ref{app:background} provides the codec lineage and technical background omitted from the main argument.

\paragraph{Contributions.}
\begin{itemize}
\item We organize audio generative modeling around the coupled design of an audio latent and the model that learns its distribution.
\item We formulate representation design as a four-objective Pareto problem over burden, distortion, empirical modelability, and streaming compatibility.
\item We use dependency horizon, conditional ambiguity, approximation sources, and critical-path evaluations as diagnostic tools for comparing autoregressive, codebook-depth, iterative, and hybrid systems, and apply them to representative systems.
\end{itemize}

\section{Two Decisions Behind Audio Generation}
\label{sec:setup}

Let $x\in\mathbb{R}^{L}$ denote a waveform of duration $\mathcal T$. An encoder maps it to $z=(z_1,\dots,z_N)$ at frame rate $r=N/\mathcal T$. For a discrete latent, each $z_t$ contains one or more codebook indices; for a continuous latent, $z_t\in\mathbb{R}^d$. We write the encoder and decoder as $f_\theta$ and $g_\phi$, and let $p_\psi$ model $p(z)$ or $p(z\mid c)$.

Two representation-level quantities can be written directly:
\begin{align}
R_{\mathrm{disc}} &= r\sum_{k=1}^{K}\log_2|\mathcal{Z}_k|, &
D &= \mathbb{E}_{x}\,\rho\big(x,g_\phi(f_\theta(x))\big). \label{eq:RD}
\end{align}
$R_{\mathrm{disc}}$ is the nominal bitrate of a $K$-codebook discrete representation, and $D$ is perceptual distortion. Rate-distortion theory motivates comparing these two quantities under a source and distortion measure~\citep{coverthomas}. A continuous latent has no codec bitrate until a precision or entropy model is specified, so discrete and continuous rates should not be collapsed into one scalar. For discrete latents, we report frame rate, codebook depth, and nominal bitrate; for continuous latents, we report frame rate, dimension, and numerical precision.

Modelability is not a third intrinsic scalar. We report it as an empirical profile indexed by task $q$, dataset $\mathcal D$, conditioning $c$, modeling scheme $G$, architecture family $A$, training budget $B_{\mathrm{train}}$, and inference budget $B_{\mathrm{infer}}$:
\begin{equation}
\mathcal M_{\mathrm{emp}}(f;q,\mathcal D,c,G,A,B_{\mathrm{train}},B_{\mathrm{infer}})
=\big(\mathcal L_{\mathrm{val}},Q_{\mathrm{gen}},C_{\mathrm{train}},C_{\mathrm{infer}}\big).
\label{eq:modelability-profile}
\end{equation}
Here $\mathcal L_{\mathrm{val}}$ is a within-objective validation loss, $Q_{\mathrm{gen}}$ contains task and perceptual generation measures, and $C_{\mathrm{train}}$ and $C_{\mathrm{infer}}$ record realized resource use. Losses from token likelihood, denoising, and flow matching are not compared numerically across objectives. A broad modelability claim should show that a ranking persists across several suitable $G$ and $A$, rather than selecting one architecture that favors a latent.

Streaming also requires a profile. Encoder lookahead can be defined as
\begin{equation}
\tau_{\mathrm{enc}}=\inf\{\,\Delta\ge 0: z_t \text{ depends only on }x_{\le t/r+\Delta}\,\},
\end{equation}
but system latency additionally includes decoder lookahead, block duration, critical-path network evaluations, computation time, and time to first audio. We therefore use streaming compatibility as a representation objective and report end-to-end latency for the complete pair.

\subsection{A Four-Objective Representation Design Space}
\label{sec:conv-tetra}

These four quantities define the design space used in the rest of the paper:
\begin{itemize}
\item \textbf{Representation burden}: frame rate and per-frame capacity.
\item \textbf{Distortion}: how much perceptual quality is lost.
\item \textbf{Empirical modelability}: the profile in Equation~\ref{eq:modelability-profile} under declared tasks and budgets.
\item \textbf{Streaming compatibility}: encoder causality and the latency induced when the representation is paired with a generator and decoder.
\end{itemize}

\fig{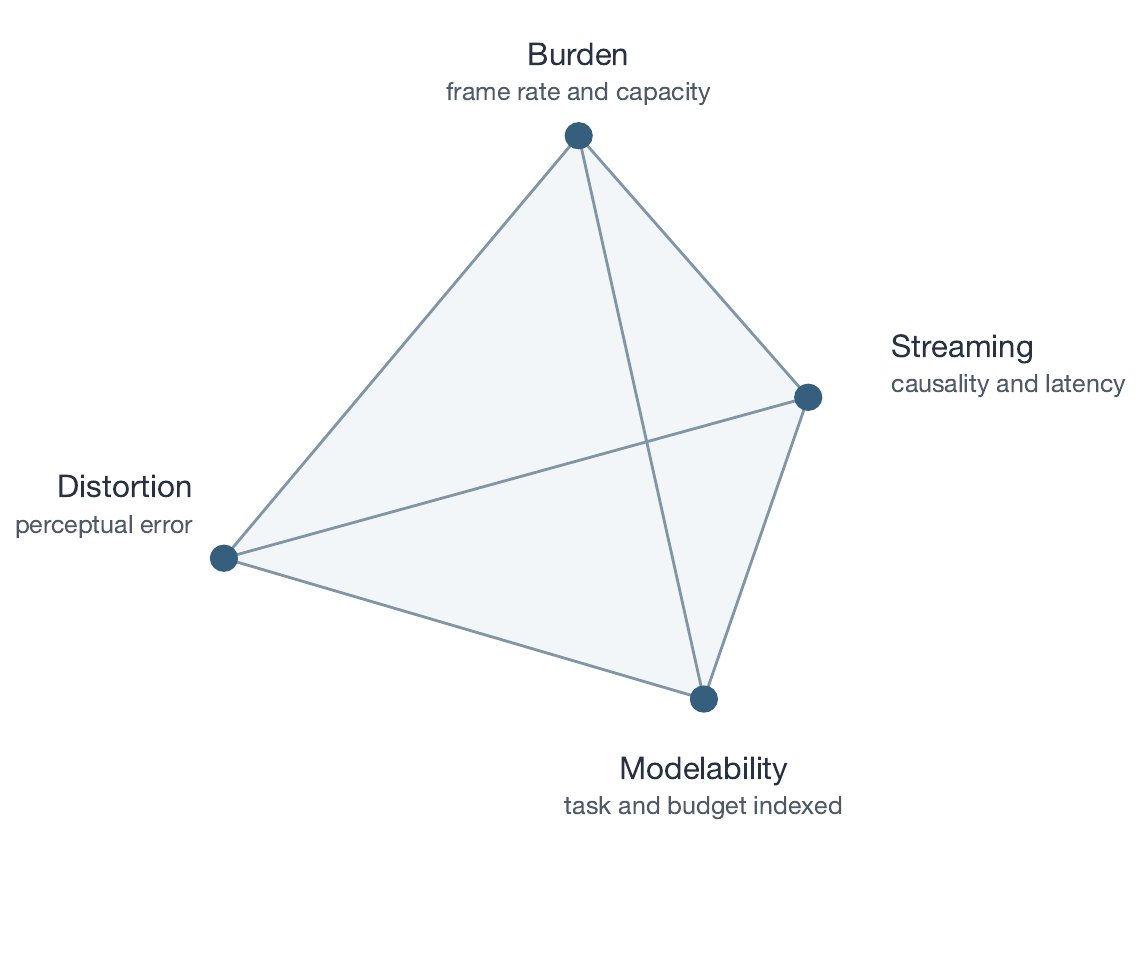}{0.4}{A schematic view of four objectives in audio representation design. The tetrahedral drawing is a mnemonic for interacting Pareto objectives, not a metric geometry: the axes have different units and are not barycentric coordinates.}

Concrete design questions couple these objectives: compression versus reconstruction quality; frame rate versus per-frame modeling burden; reconstruction-optimal versus generator-readable latents; fixed-rate versus content-adaptive streaming; and offline quality versus causal generation. The four values do not define a common numerical coordinate system. The figure is a design mnemonic. Empirical results should instead be reported as a Pareto set. Discrete and continuous representations are implementation choices within this space, not competing vertices.

Compatibility with a pretrained text LLM is not a fifth intrinsic axis. It is modelability relative to a particular model family $A$. Discrete tokens fit its softmax interface but can dominate context length and need not align with text semantics. LLM-Codec~\citep{llmcodec} addresses this directly by deriving its codebook from a pretrained LLM's token embeddings, so that audio tokens occupy the same vocabulary space the language model already uses. Continuous latents require projectors and a generative output head but may align more naturally with hidden states. The relevant comparison fixes the shared backbone and compute budget.

\subsection{When Is a Representation Easier to Model?}

What does it mean to say that a latent is ``easy to model''? It cannot mean that the latent carries an intrinsic difficulty score. Modelability depends on the task, dataset, conditioning, modeling scheme, architecture family, and both training and inference budgets. The same RVQ stream may be difficult for a flat transformer but easier for a global-over-time, local-over-depth architecture; the same continuous latent may fail under deterministic regression but work under a conditional flow head. A narrow comparison should match every field in Equation~\ref{eq:modelability-profile}. A broader claim should evaluate several suitable model families and report the resulting Pareto sets.

In practice, four measurements expose different failure modes:
\begin{enumerate}
\item \textbf{Sequence burden}: frames or tokens per second, including the effective expansion caused by multiple codebooks.
\item \textbf{Predictive burden}: held-out loss under the same objective, latent normalization, model family, and training budget.
\item \textbf{Scaling burden}: quality as parameters, context length, training FLOPs, or sampling steps increase.
\item \textbf{Task burden}: semantic stability, speaker and prosody retention, and perceptual quality in generated samples, not only codec reconstruction.
\end{enumerate}
No single one is sufficient. Loss values from different objectives or latent coordinate systems are not directly comparable. Low token entropy can be manufactured by collapsing useful information. Good reconstruction can coexist with an irregular latent distribution. Low training loss can reflect an easy but semantically empty target. The meaningful comparison is downstream quality at matched reconstruction quality, latency, and compute.

\begin{principle}[Reconstruction does not determine modelability]
\label{prin:ib}
A reconstruction-only codec solves
\begin{equation}
\begin{aligned}
(\theta_{\mathrm{rec}},\phi_{\mathrm{rec}})
&=\arg\min_{\theta,\phi} \\
&\quad \mathbb E_x\!\left[\rho\!\left(x,g_\phi(f_\theta(x))\right)\right].
\end{aligned}
\label{eq:recon-only}
\end{equation}
A representation designed together with its generator instead solves an objective of the form
\begin{equation}
(\theta^\star,\phi^\star,\psi^\star)
=\arg\min_{\theta,\phi,\psi}
\mathbb E_{x,c}\!\left[
\rho\!\left(x,g_\phi(f_\theta(x))\right)
+\lambda\,\mathcal L_{\mathrm{gen}}\!\left(f_\theta(x);p_\psi(\cdot\mid c)\right)
\right].
\label{eq:joint-objective}
\end{equation}
The first objective only requires the latent to work with its decoder. The second also requires it to work with the downstream generator. Their optima need not coincide because reconstruction quality does not specify how the latent distribution is organized for a finite generator. The information bottleneck provides a related view: a useful representation should preserve task-relevant information and discard irrelevant variability~\citep{ib}.
\end{principle}

DAC~\citep{dac} documents this gap concretely. Its objective combines adversarial and perceptual losses without a dominant sample-level reconstruction term. Its reconstructions can therefore score poorly on waveform-level fidelity measures such as SI-SNR but score strongly on perceptual measures such as ViSQOL~\citep{survey}. The decoder resynthesizes a perceptually equivalent waveform rather than one that matches the reference sample by sample. This is a concrete instance of Principle~\ref{prin:ib}: a distortion measure evaluated on samples and one evaluated by a listener or a downstream task need not agree, and a latent judged only by the former has not yet been judged by the criterion a downstream system actually needs.

\subsection{How Should a Latent's Distribution Be Modeled?}

The representation alone does not decide how its distribution should be modeled. The same latent can be factorized in different orders, or have its remaining uncertainty represented by different generative mechanisms. Which choice is appropriate depends on how far its dependencies extend and how much ambiguity is left after conditioning.

\begin{principle}[Predict long-range structure, generate ambiguous detail]
\label{prin:entropy}
For an autoregressive model,
\begin{equation}
H(z)=\sum_{t=1}^N H(z_t\mid z_{<t}).
\label{eq:chain}
\end{equation}
For a discrete autoregressive target, the irreducible next-step log loss is the conditional entropy. Let $s_t$ denote structure that may require long-context prediction, and let $a_t$ denote the remaining acoustic realization, so that $z_t=(s_t,a_t)$. The chain rule gives
\begin{equation}
H(s_t,a_t\mid z_{<t})=
H(s_t\mid z_{<t})+H(a_t\mid z_{<t},s_t).
\label{eq:split}
\end{equation}
This decomposition is often summarized as low-entropy semantics followed by high-entropy acoustics. The intuition is useful: content is constrained by linguistic or musical structure. The same content can still admit many perceptually valid acoustic realizations. But the two marginal entropies do not by themselves determine an architecture. Semantic continuation can remain highly uncertain, and speaker identity, prosody, or musical form can give acoustics long-range structure. Our design hypothesis is more specific: $s_t$ should expose the uncertain variables whose dependencies reward long-context prediction. $H(a_t\mid z_{<t},s_t)$ measures the remaining realization ambiguity that a conditional acoustic generator must represent. For continuous latents, $s_t$ and $a_t$ shape the conditional distribution instead of a token entropy that can be compared numerically.
\end{principle}

This is the lens we use throughout the paper, but $s_t$ and $a_t$ need not be explicit coordinates of a learned latent. They denote functional roles that may remain entangled in practice. We treat dependency horizon and conditional ambiguity as diagnostic dimensions, not universal latent statistics.

A context-ablation curve gives one proxy for dependency horizon. For a fixed representation, task, model family, and budget, let $\mathcal L(h)$ be validation loss when the usable history is truncated to duration $h$, and define
\begin{equation}
h_\delta=\min\{h:\mathcal L(h)\leq \mathcal L(h_{\max})+\delta\}.
\label{eq:horizon-proxy}
\end{equation}
The curve, rather than $h_\delta$ alone, shows whether additional history remains useful. Conditional ambiguity is relative to a conditioning set and a perceptual equivalence criterion. Practical proxies include diversity among perceptually acceptable samples at fixed conditioning and the quality gap between a deterministic predictor and a matched distributional head. Neither proxy identifies a unique semantic-versus-acoustic split, but together they turn the two concepts into testable questions.

\section{Why Audio Representation Design Is Hard}
\label{sec:audio}

Audio representation design is constrained by three properties of the signal and the generation task.

\textbf{Long temporal unrolling.} Audio duration translates directly into sequence length. Even after aggressive downsampling, a 30-second sample can contain hundreds or thousands of latent steps. Frame rate therefore controls context length, serial generation cost, and the amount of state that a streaming model must maintain. This is the physical source of the representation burden objective in Section~\ref{sec:conv-tetra}: audio duration forces frame rate to be actively managed rather than a free parameter.

\textbf{Perceptual ambiguity.} Pointwise waveform agreement is not the same as perceptual agreement. Two signals can remain perceptually equivalent even when phase, noise texture, reverberation, and micro-timing differ. Mel-spectrogram systems exploited this freedom by leaving waveform realization to a vocoder, and learned codecs such as DAC~\citep{dac} can reconstruct natural audio without matching the waveform sample by sample. Audio latents should therefore avoid forcing the long-context model to resolve every perceptually valid local alternative. This is the physical source of the conditional ambiguity term $a_t$ in Principle~\ref{prin:entropy}: much of what a codec discards is not lost information but detail with more than one acceptable realization.

\textbf{Structure unfolds over time.} Linguistic content, melody, rhythm, speaker state, and prosody evolve at different temporal scales. Locally plausible acoustic frames do not guarantee that these trajectories remain coherent. A useful latent must keep the long-range variables accessible to the model responsible for planning them. This is the physical source of dependency horizon and the term $s_t$ in Principle~\ref{prin:entropy}: these are exactly the variables whose prediction rewards long context.

The mel-spectrogram already embodied part of this design before learned codecs. It discarded phase, retained a lower-rate time-frequency scaffold, and left waveform realization to a vocoder. Learned codecs and continuous audio latents retain the same broad division. They learn what the scaffold should preserve instead of fixing it by design.

A useful audio latent should keep what the downstream model needs to plan over time, including content, rhythm, speaker identity, and coarse acoustic structure. It need not expose phase and fine waveform texture to the long-context generator when a decoder or local conditional model can realize them.

That is the common thread behind mel-spectrograms, semantic tokens, RVQ layers, and continuous SSL-based latents. Each one shortens the waveform and reshapes it into something the generator can model without giving up the information needed to recover natural audio.

\section{Discrete Audio Latents}
\label{sec:discrete}

Discrete latents turn audio generation into categorical prediction, but discreteness alone does not make audio language-like. The downstream burden is determined by four choices: how often a token is emitted, how much information each frame carries, whether content is visible in the token stream, and how much realization ambiguity remains once that content is fixed. We refer to these as temporal rate, per-frame capacity, dependency horizon, and conditional ambiguity. The first two instantiate representation burden (Section~\ref{sec:conv-tetra}); the last two are Principle~\ref{prin:entropy}'s pair of diagnostic dimensions, applied to a discrete latent's codebook structure.

For a $K$-codebook representation, the relevant question is where the dependencies are placed. Its entropy can be expanded along time and codebook depth as
\begin{equation}
H(z)=\sum_{t=1}^{N}\sum_{k=1}^{K}
H\!\left(z_t^k\mid z_{<t}^{1:K},z_t^{<k}\right).
\label{eq:rvq-chain}
\end{equation}
The equality itself favors no architecture. Under a finite budget, however, the placement matters. Frame rate determines how many temporal decisions the model must make. Codebook depth determines how many conditional decisions remain inside each frame. Dependency horizon determines whether long-range content can be predicted directly or must be inferred through acoustically organized codes.

\subsection{Temporal rate and per-frame capacity}
\label{sec:rvq}

A single codebook gives one categorical decision per frame and fits a standard LM interface. At a fixed codebook size, preserving more acoustic detail usually requires a higher frame rate. RVQ takes the other route: it keeps the temporal rate low and adds capacity in codebook depth. The first quantizer represents a coarse approximation, and later quantizers encode successive residuals.

\fig{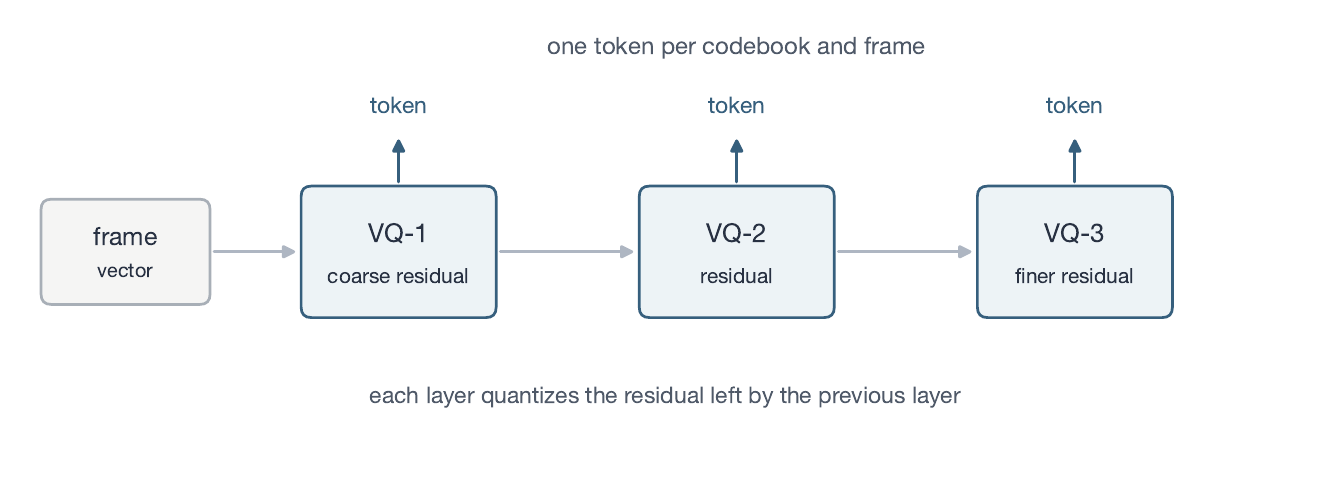}{1.0}{RVQ's bet: build capacity in depth, not by raising the frame rate. The first quantizer is coarsest; each later layer quantizes the previous residual. One frame can therefore hold enough information for fidelity even at a low frame rate.}

If the frame rate is $r$ and there are $K$ codebooks, flattening gives $Kr$ token decisions per second. Parallel prediction keeps the temporal length at $r$, but each frame becomes a structured $K$-variable problem. RVQ therefore does not remove modeling cost. It moves capacity from temporal length into depth, where a downstream architecture may exploit a cheaper local factorization. This is a potential modeling advantage, alongside RVQ's primary codec benefit of scalable bitrate and fidelity.

Representative systems make the trade-off visible. SoundStream~\citep{soundstream} and EnCodec~\citep{encodec} established moderate-frame-rate, multi-codebook RVQ streams. HiFi-Codec~\citep{hificodec} and DAC~\citep{dac} changed the quantizer or reconstruction design within the same broad template. Mimi~\citep{moshi} pushes the frame rate to 12.5~Hz for real-time dialogue by accepting substantial per-frame depth. In contrast, WavTokenizer~\citep{wavtokenizer} and BigCodec~\citep{bigcodec} pursue a single-codebook interface at roughly 75--80~Hz. This is not a direct leaderboard comparison because training data, bandwidth, causality, and evaluation differ. It nevertheless exposes the basic exchange: fewer time steps require more information per step.

This comparison explains why RVQ remains difficult to replace at the codec layer. SVQ offers a simpler interface, but often pays through a higher temporal rate or a larger token space. FSQ-style designs~\citep{fsq} improve code utilization, but their scalar dimensions do not automatically provide an ordered prediction schedule. RVQ provides ordered capacity. It does not provide ordered semantics. The same rate-versus-depth exchange reappears for continuous latents as a rate-versus-dimension exchange (Section~\ref{sec:continuous}). RVQ's specific mechanism is not what makes representation burden a real constraint.

This capacity-versus-rate trade-off is not resolved by the codec. It is inherited by whatever model consumes the resulting $K$-codebook stream. A flattened autoregressive model pays the full $NK$ serial cost implied by Equation~\ref{eq:rvq-chain} to keep every conditional dependency available. VALL-E~\citep{valle} instead spends serial computation only on the first layer and predicts the remaining $K-1$ layers non-autoregressively, trading some cross-layer conditioning for a shorter critical path. MusicGen~\citep{musicgen} keeps a causal ordering across codebooks but staggers them with a delay pattern, so several codebooks can be predicted in parallel without discarding the ordering entirely. UniAudio~\citep{uniaudio} and Moshi~\citep{moshi} instead factor the problem explicitly: a temporal transformer models frame-to-frame dependence, and a smaller local transformer models the codebook-depth dependence within each frame. SoundStorm~\citep{soundstorm} and MAGNeT~\citep{magnet} replace the causal order with iterative masked refinement across the token grid. Each scheme keeps Equation~\ref{eq:rvq-chain}'s dependencies exact or approximates a specific subset of them. Section~\ref{sec:modeling} and Appendix~\ref{app:downstream} compare their critical-path cost directly.

\subsection{Dependency Horizon in a Discrete Latent}
\label{sec:audiolm}

A concrete failure motivates this section. Early autoregressive models trained on SoundStream~\citep{soundstream} tokens alone could preserve locally plausible speech but lose linguistic coherence during continuation: content could drift even though pronunciation remained speech-like. Codec reconstruction could not reveal this failure, since the decoder was given ground-truth tokens. Generation exposed it, because one acoustically organized stream asked the long-context model to maintain content and predict local realization detail at once.

This is a specific instance of a general requirement. A reconstruction codec only needs its decoder to recover audio from $z$. A generative model needs a stronger property, per Principle~\ref{prin:ib}: task-relevant long-range information must be accessible to the model that predicts $z$, not only to its decoder. Standard RVQ does not guarantee this: its early layers contain whatever most reduces reconstruction error, which may mix phonetic content, pitch, speaker, and channel information.

AudioLM~\citep{audiolm} is the system that exposed this failure, and its fix illustrates one way to guarantee accessibility rather than a template every system must follow. It added semantic tokens from w2v-BERT~\citep{w2vbert}: a semantic LM predicted a low-rate content trajectory, and acoustic LMs generated SoundStream tokens conditioned on it. In terms of Principle~\ref{prin:entropy}, the system placed long-range content dependence in the sequence-level variable $s$, the aggregate of $s_t$ across time, and left conditional acoustic variation in $p(a\mid s)$, where $a$ is the corresponding aggregate of $a_t$; Section~\ref{sec:discrete-ambiguity} addresses how that second term is handled.

\fig{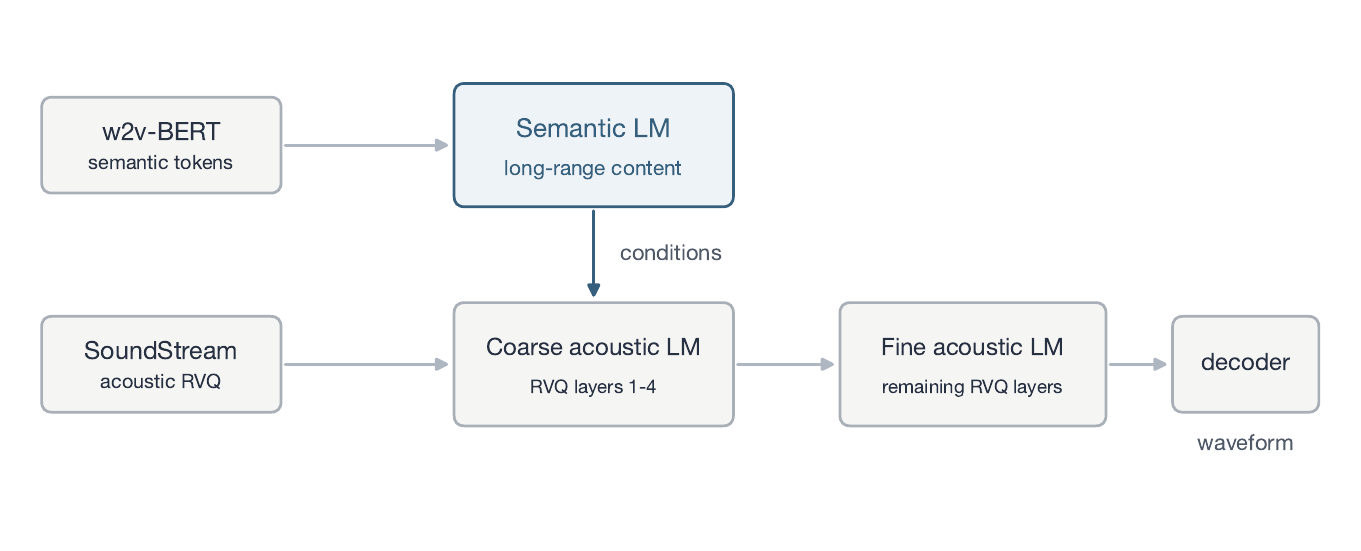}{0.65}{AudioLM's cascade: semantic tokens anchor long-range content. Acoustic tokens recover timbre, prosody, and detail. The important idea is the division of labor, not the exact three-stage implementation.}

Semantic supervision changes where information appears in Equation~\ref{eq:rvq-chain}. SpeechTokenizer~\citep{speechtokenizer}, SoCodec~\citep{socodec}, Mimi~\citep{moshi}, and SemantiCodec~\citep{semanticodec} align selected codec components with SSL features or semantic targets. X-Codec~\citep{xcodec} combines semantic and acoustic encoder features before quantization. HuBERT~\citep{hubert}, WavLM~\citep{wavlm}, or ASR targets can provide the supervisory signal. MOSS-Audio-Tokenizer~\citep{mossaudiotokenizer} instead uses LM-based understanding tasks to strengthen semantic content. These methods do not merely improve a token representation. They try to place content in a low-rate component that a long-context model can read directly. Semantic supervision is one way to satisfy Principle~\ref{prin:ib}, not the only conceivable one: the principle only asks that task-relevant information be accessible to the predicting model, not that an external semantic signal was used to place it there. Whether a reconstruction-only discrete latent can reach comparable accessibility through favorable geometry or training dynamics alone remains an open question, addressed further in Section~\ref{sec:open}.

This need is strongest when the model must continue without a complete external content plan. In TTS, text already constrains content, so acoustic-token-only systems can work. Even there, predicting a compact semantic or phonetic plan first usually simplifies the acoustic mapping. The difference is one of degree, not a binary distinction between continuation and TTS.

AudioLM also makes an assumption that should remain visible: once semantics is known, most remaining acoustic dependence is local enough for a cheaper model. Speaker identity, expressive prosody, room acoustics, and musical form can violate that assumption. The useful principle is therefore not ``semantics first'' in every system. It is to expose each long-range factor before asking a local generator to fill in detail.

Dependency horizon must be balanced with reconstruction and rate under a finite latent capacity. Adding semantic supervision is useful only if it reduces downstream modeling difficulty at matched fidelity and latency. This is why generator-aware objectives, as in ALMTokenizer~\citep{almtokenizer}, are conceptually important: they optimize the interface with the downstream model rather than treating reconstruction as the final criterion. Continuous latents face the same accessibility problem without a codebook layer to supervise; Section~\ref{sec:continuous} addresses how they shape a comparably accessible geometry instead.

Modern systems preserve this division in three ways. Layer assignment places semantic supervision in selected codec layers, as in SpeechTokenizer~\citep{speechtokenizer} and Mimi~\citep{moshi}. Factor assignment separates time-invariant information in TiCodec~\citep{ticodec}, sources in SD-Codec~\citep{sdcodec}, or content, prosody, speaker, and residual variables in FACodec~\citep{ns3}. Model assignment predicts semantic or phonetic variables first and uses a conditional acoustic generator afterward, as in CosyVoice~\citep{cosyvoice} and Seed-TTS~\citep{seedtts}. The first two reshape the representation itself. The third leaves the representation alone and reallocates the work to the modeling side instead, a preview of the coupling explored in Section~\ref{sec:modeling}.

\subsection{Conditional Ambiguity in a Discrete Latent}
\label{sec:discrete-ambiguity}

Exposing the long-range content $s$ does not remove the conditional acoustic detail $a$ left once $s$ is known. RVQ's later codebook layers still carry pitch micro-detail, timbre, phase-related texture, and other realization variables for which more than one value is perceptually acceptable given the same content. This is the discrete-latent instance of conditional ambiguity in Principle~\ref{prin:entropy}.

Categorical prediction has a property continuous regression lacks: a softmax over a codebook is already a distribution, and sampling from it already represents more than one acceptable realization. Unlike a continuous latent, where a deterministic regressor collapses ambiguity toward a conditional mean and a distributional mechanism must be added deliberately (Section~\ref{sec:cont-loss}), a discrete latent gets a basic notion of conditional variation for free from its output layer.

This does not close the question, because Equation~\ref{eq:rvq-chain}'s chain rule still specifies exactly which variables each layer may condition on, and no practical model implements it exactly. VALL-E~\citep{valle} predicts later layers non-autoregressively, approximately treating them as conditionally independent given the first layer. This is fast, but it can discard correlations between codebooks that a truly joint sample would preserve. SoundStorm~\citep{soundstorm} and MAGNeT~\citep{magnet} instead resolve the remaining tokens through iterative masked refinement, revisiting the same positions multiple times so that later predictions can depend on tokens filled in earlier in the schedule rather than committing to full conditional independence in one pass. AudioLM's~\citep{audiolm} coarse-to-fine acoustic LMs are a third compromise: each stage conditions on the previous stage's output, narrowing the realization progressively instead of resolving it in one non-autoregressive step or many masked rounds.

None of these schemes is uniquely correct. They trade the same two costs already introduced in Section~\ref{sec:rvq}: how much of Equation~\ref{eq:rvq-chain}'s conditional structure is preserved, and how many sequential network evaluations that preservation costs. Section~\ref{sec:modeling} and Appendix~\ref{app:downstream} compare these schedules directly on that basis.

Taken together, Sections~\ref{sec:rvq}--\ref{sec:discrete-ambiguity} point to three levers the literature has used to shape a discrete latent's modelability, none of them guaranteed to help outside the setting in which it was demonstrated.
\begin{itemize}
\item \textbf{Representation burden} can be moved between frame rate and codebook depth (RVQ) or concentrated into a single larger vocabulary (SVQ, FSQ), changing how many temporal versus conditional decisions a downstream model faces.
\item \textbf{Dependency horizon} can be made accessible by distilling SSL or ASR targets into selected codec components, by training the tokenizer against a generator-aware objective instead of reconstruction alone, or by explicitly factoring content, prosody, and speaker into separate variables.
\item \textbf{Conditional ambiguity}, once content is exposed, is preserved to a degree that depends on whether the model approximates Equation~\ref{eq:rvq-chain}'s chain rule by conditional independence, iterative masked refinement, or a staged cascade.
\end{itemize}
None of these choices is free: each trades reconstruction, rate, or critical-path cost for the property it improves, which is why Section~\ref{sec:open} asks for controlled comparisons rather than treating any one of them as a default recipe.

\section{Continuous Audio Latents}
\label{sec:continuous}

Continuous latents remove categorical quantization, but they do not remove the representation problem. Their downstream burden is controlled by the same four choices as a discrete latent: temporal rate, per-token dimension, dependency horizon, and conditional ambiguity. The first two instantiate representation burden (Section~\ref{sec:conv-tetra}). The last two are Principle~\ref{prin:entropy}'s diagnostic dimensions, applied to a continuous latent's geometry instead of its codebook structure.

Let $y$ denote structure that the downstream system must preserve, such as linguistic content, speaker identity, or musical form. Two information terms clarify the tension. $I(z;y)$ measures how much of that structure the latent retains. $I(z;x\mid y)$ measures how much additional acoustic variation it carries once that structure is known. $y$ plays the role of dependency horizon's $s_t$ from Principle~\ref{prin:entropy} for a continuous latent, and $I(z;x\mid y)$ is its conditional ambiguity term $a_t$, expressed as retained information rather than token entropy. Retention does not guarantee accessibility to a finite model, so semantic probes or matched downstream losses are still needed. The second term is not simply waste: some of it is required for fidelity. But if a high-dimensional latent exposes every local variation to the generator, the downstream model must fit that variation even when it is perceptually ambiguous. Continuous latent design decides how much of this conditional detail stays in $z$ and how much is left to the decoder.

\subsection{Temporal rate and per-token dimension}
\label{sec:cont-burden}

A continuous latent trades frame rate against per-token dimension the same way a discrete latent trades frame rate against codebook depth: more information per step allows a lower rate, and a lower rate keeps the downstream sequence short. Stable Audio~\citep{stableaudio}'s VAE, in one released configuration, compresses 44.1~kHz stereo audio by a factor of 2048, giving a 64-dimensional latent at roughly 21.5~Hz. NaturalSpeech~2~\citep{naturalspeech2} instead keeps a much higher frame rate, close to 80~Hz, with a single continuous vector per frame rather than a multi-codebook stream. CALM~\citep{calm} uses a 128-dimensional audio-VAE frame in at least one open implementation. These are not a leaderboard comparison, since the encoders, domains, and training data differ. They show the same exchange as Section~\ref{sec:rvq}'s RVQ comparison, expressed in dimension rather than codebook depth. Fidelity can be bought with a higher frame rate and low per-token dimension, or with a lower frame rate and high per-token dimension. The choice determines how many autoregressive or diffusion steps the downstream model must take, not only how the encoder compresses audio.

\subsection{Dependency Horizon in a Continuous Latent}
\label{sec:cont-semantic}

Mel-spectrograms and VAE latents expose a compact acoustic scaffold and leave waveform detail to a decoder. Make-An-Audio~\citep{makeanaudio}, NaturalSpeech 2~\citep{naturalspeech2}, and Stable Audio~\citep{stableaudio} demonstrate this route across general audio, speech, singing, and music. Such latents are efficient generation targets, but reconstruction training does not require their coordinates to expose content. SSL features take the opposite starting point. They make phonetic or semantic structure accessible, but are often high-dimensional and retain variations that a generator must reproduce at every frame.

Recent work tries to build a compressed semantic-acoustic latent between these endpoints. DashengTokenizer~\citep{dasheng} starts from frozen semantic features and injects acoustic information. WavCube~\citep{wavcube} compresses an SSL representation before acoustic reconstruction and retains a semantic anchor throughout. MingTok-Audio~\citep{minguniaudio} and LoSATok~\citep{losatok} likewise seek representations shared by understanding and generation.

In information terms, semantic alignment aims to preserve information about $y$ in a form accessible to the selected model. Bottlenecking limits the conditional detail exposed to the generator. Reconstruction prevents the latent from discarding information required by the decoder. These operations serve different purposes. A high-dimensional semantic latent is not automatically modelable, and a low-dimensional latent is not automatically well organized.

\subsection{Conditional Ambiguity in a Continuous Latent}
\label{sec:cont-loss}

Continuous dimension is not a bitrate, but it directly affects the output problem. The failure mode is not hypothetical: early neural TTS systems that predicted mel-spectrograms under a deterministic $L_1$ or $L_2$ loss produced characteristically over-smoothed, muffled spectral detail, because squared error averaged over every acoustically valid realization of a given phoneme and prosody collapses toward their mean rather than committing to any one of them. A deterministic regressor must choose one point in latent space. When several acoustic realizations are valid, squared error moves the prediction toward their conditional mean. Diffusion, flow matching~\citep{flowmatching}, mixture heads, and other distributional objectives are useful because they can represent this conditional variation. Voicebox~\citep{voicebox} is an influential example of using flow matching for conditional speech infilling and generation.

These mechanisms are not interchangeable, and audio's own modeling literature has not worked out their trade-offs as thoroughly as image generation has. GIVT~\citep{givt} replaces the categorical output of a discrete-token transformer with a Gaussian mixture head, predicting a full per-token distribution in one forward pass but limiting expressivity to the chosen mixture family. MAR~\citep{mar} instead attaches a small diffusion head to each token, trading a fixed per-token sampling cost of several denoising steps for a distribution that is not constrained to a parametric family. CALM~\citep{calm} takes a related route for audio, distilling a consistency model so that most of diffusion's per-step sampling cost is paid once, offline, rather than at every generation step. The choice is the same trade-off already introduced for discrete latents in Section~\ref{sec:discrete-ambiguity}: how much of the true conditional distribution a mechanism can represent, against how many network evaluations it costs to sample from it.

Continuous geometry matters even when reconstruction does not change. For any invertible transform $T$, define $z'=T(z)$ and $g'_\phi(z')=g_\phi(T^{-1}(z'))$. The transformed pair preserves the same reconstruction,
\begin{equation}
g'_\phi(z')=g_\phi(z),
\end{equation}
but it presents a different density geometry to a finite downstream model. Reconstruction therefore cannot select among all equivalent coordinate systems. This is Principle~\ref{prin:ib} applied to continuous geometry specifically: two encoder-decoder pairs can satisfy the same reconstruction objective but present entirely different learning problems to a finite generator. If semantic and nuisance variation are entangled across many dimensions, reducing frame rate alone may still leave a difficult per-step distribution. Continuous latent design must be evaluated together with the downstream loss and model family, just as RVQ must be evaluated together with its depth factorization.

A convergent observation appears in continuous-token autoregressive image generation. Constraining VAE latents to a fixed-radius hypersphere removes a scale degree of freedom that otherwise drives variance collapse under classifier-free guidance. This geometry change alone lets a purely autoregressive model match or exceed diffusion and masked-generation baselines at comparable scale~\citep{spherear}. The reconstruction target is unchanged; only the coordinate geometry presented to the autoregressive model is.

High-dimensional SSL features often expose long-range structure well but create a large per-frame prediction problem. Low-dimensional VAE latents reduce that burden but may hide the same structure. Compressed semantic latents try to preserve accessibility and still control dimension and conditional acoustic variation. They remain subject to the four-objective design space: compression can increase distortion, semantic alignment can raise representation burden, and non-causal encoders can improve empirical modelability at the expense of streaming compatibility.

Taken together, Sections~\ref{sec:cont-burden}--\ref{sec:cont-loss} point to a comparable set of levers for a continuous latent's modelability, none of them guaranteed to help outside the setting in which it was demonstrated.
\begin{itemize}
\item \textbf{Representation burden} can be moved between frame rate and per-token dimension (Section~\ref{sec:cont-burden}), the continuous analogue of RVQ's rate-versus-depth exchange.
\item \textbf{Dependency horizon} is harder to guarantee than in a discrete latent, because there is no codebook layer to supervise directly. Recent tokenizers instead start from a frozen or lightly adapted semantic encoder and inject acoustic detail afterward, rather than the reverse: DashengTokenizer~\citep{dasheng} freezes a pretrained semantic encoder and adds acoustic information through a single linear projection from mel-spectrogram features; WavCube~\citep{wavcube} first trains a semantic bottleneck to remove the off-manifold redundancy that makes raw SSL features difficult to generate from, then injects fine-grained acoustic detail under a semantic-anchoring loss that keeps the representation from drifting off that manifold; LoSATok~\citep{losatok} compresses a high-dimensional semantic encoder's features into a much lower-dimensional space under dual-level semantic supervision. This trades dimension for accessibility but keeps enough capacity for acoustic detail.
\item \textbf{Conditional ambiguity}, once content is exposed, is preserved to a degree that depends on the same choice as in a discrete latent, but expressed through a distributional output head instead of a chain rule: a Gaussian mixture, a diffusion or consistency process, or a deterministic regressor that discards it entirely.
\end{itemize}
None of these choices is free: compressing toward a semantic encoder can raise distortion or discard acoustic detail the decoder still needs, which is why Section~\ref{sec:open} again asks for controlled comparisons rather than a default recipe.

How these latents are generated is a separate question, and we return to it in the next section.

\section{How to Model an Audio Latent}
\label{sec:modeling}

Choosing a latent defines the random variables and exposes a particular dependency structure. The modeling architecture must decide which dependencies receive causal context, which are handled locally, and which variables can be refined jointly. For a $K$-stream representation, let $\pi$ be an ordering of its $NK$ variables. A sequential model can use the exact chain-rule factorization
\begin{equation}
p(z\mid c)=\prod_{i=1}^{NK}
p\!\left(z_{\pi_i}\mid z_{\pi_{<i}},c\right).
\end{equation}
Practical architectures choose the order $\pi$, restrict the available context, or predict groups of variables together. Iterative methods make a different choice. Masked models, following the broader masked-generation idea exemplified by MaskGIT~\citep{maskgit}, learn repeated conditional refinement. Diffusion~\citep{ddpm} and flow matching~\citep{flowmatching} instead learn a corruption reversal or transport process over groups of variables. They are modeling schemes, but not alternative left-to-right chain-rule orderings.

Their practical difference appears under finite data, capacity, optimization, and generation time. Within one declared training objective, a conceptual decomposition is
\begin{equation}
\mathcal L_{\mathrm{val}}=
\mathcal L_G^\star+\epsilon_{\mathrm{fact}}+\epsilon_{\mathrm{approx}}
+\epsilon_{\mathrm{est}}+\epsilon_{\mathrm{opt}},
\label{eq:factor-gap}
\end{equation}
where $\mathcal L_G^\star$ is the irreducible risk for scheme $G$, and the four residuals denote restrictions from the chosen factorization or context, finite architecture, finite data, and imperfect optimization. Finite-step samplers add inference truncation error that does not appear in training loss. This decomposition is diagnostic rather than identifiable: ordinary experiments cannot estimate its terms separately. Exact discrete likelihood has $\mathcal L_G^\star=H(z\mid c)$. Denoising and flow objectives instead define different risks and must not be compared numerically with token entropy.

Let $E_G$ denote critical-path network evaluations: the network calls that must occur sequentially before an output segment is available. $E_G$ is more precise than token count but is still not latency. It must be reported with FLOPs, hardware, real-time factor, and time to first audio. A practical scheme must control validation and generation quality under these realized costs.

Three diagnostic dimensions guide this choice. \emph{Dependency horizon} asks how model performance changes as usable context grows. \emph{Conditional ambiguity} asks whether multiple perceptually acceptable realizations remain under a declared conditioning set and equivalence criterion. These questions refine the low-entropy-versus-high-entropy intuition in Principle~\ref{prin:entropy}. They ask not only how much uncertainty remains but also how it is structured. \emph{Streaming requirement} asks whether future context is available and how much end-to-end latency is allowed. Together, these dimensions suggest which modeling strategies should be tested. They do not determine one architecture from the latent alone.

\subsection{Long-range causal dependence favors autoregression}

Autoregression gives each step access to the generated past. It is valuable when causal history contains information that a parallel predictor cannot recover from conditioning alone. GSLM~\citep{gslm} and TWIST~\citep{twist} model long-range spoken content with semantic units. MusicLM~\citep{musiclm} uses semantic and acoustic hierarchies for music. Dialogue state and persistent speaker or prosodic state can have the same long dependency horizon. A causal AR factorization is compatible with streaming when the encoder, conditioning path, decoder, and computation are also causal and fast enough. Tokenwise AR has $E_G$ on the order of $N$. Block prediction can shorten this path.

This explains why low-rate semantic tokens are natural AR targets: each serial step advances a long-range state. Applying the same factorization to a high-rate acoustic stream spends serial computation on phase, texture, and other locally ambiguous details. The issue is not that AR cannot model acoustics. It is whether causal history reduces enough uncertainty to justify the serial path.

RVQ adds a second dependency horizon across codebook depth. Flattening time and depth gives rich context but can raise the critical path toward $NK$ evaluations. Section~\ref{sec:rvq} already introduced how VALL-E~\citep{valle}, MusicGen~\citep{musicgen}, and Moshi~\citep{moshi} trade retained conditioning against critical-path cost when they resolve this second horizon. The point that matters here is that the trade-off is a property of the dependency structure itself, not of any one system's specific schedule.

\subsection{Strong global conditioning favors iterative refinement}

When text, duration, or another condition strongly constrains global structure, bidirectional refinement can shorten the critical path without discarding much useful context. SoundStorm~\citep{soundstorm} and MAGNeT~\citep{magnet} repeatedly fill or revise masked codec tokens. DiffSound~\citep{diffsound} and InstructTTS~\citep{instructtts} use discrete diffusion to refine VQ acoustic tokens. Continuous diffusion and flow matching learn different mathematical objects, but make the same systems-level exchange: they update many time positions in parallel over several network evaluations.

This strategy is well matched to strongly conditioned offline TTS or text-to-audio generation. E2-TTS~\citep{e2tts} and F5-TTS~\citep{f5tts} illustrate how full conditioning and non-causal generation can support flow-based sequence synthesis. It is less natural when the system must emit audio before future conditioning or future latent positions are available. Iterative refinement is therefore selected by available context, not by whether the latent is discrete or continuous.

\subsection{Hybrids: sequential outside, iterative inside}
\label{sec:hybrids}

Many audio latents contain both long-range causal state and locally ambiguous realization. A hybrid factorization assigns them to different scales. Continuous autoregression keeps a causal outer model over time and uses a distributional head for the next frame or block. MELLE~\citep{melle}, DiTAR~\citep{ditar}, VibeVoice~\citep{vibevoice}, and CALM~\citep{calm} explore versions of this pattern. Block-causal diffusion and flow are serial across blocks and iterative within each block.

The design variables are block size and inner sampling cost. In a streaming system, larger blocks reduce outer critical-path evaluations but add lookahead and make each conditional distribution harder. More inner refinement steps can reduce local approximation error but increase latency. Calling the whole system AR, diffusion, or flow hides this allocation.

\subsection{Representative latent-model pairings}

As in Section~\ref{sec:audiolm}, let $s$ denote exposed long-range structure and $a$ the remaining acoustic realization. The chain rule gives
\begin{equation}
p(s,a)=p(s)\,p(a\mid s).
\end{equation}
The value of this order is not that semantics is always low entropy. A compact $s$ shortens the expensive serial path and still exposes long-range constraints. Conditioning on $s$ then reduces the dependency horizon of much of $a$, allowing a local or iterative acoustic model to absorb the remaining ambiguity.

AudioLM~\citep{audiolm} places the split between semantic and acoustic tokenizers. CosyVoice~\citep{cosyvoice} places it between a semantic-token LM and a continuous flow decoder. Mimi within Moshi~\citep{moshi} distills semantic information into a causal codec layer. Moshi separates temporal prediction from depth prediction. VALL-E~\citep{valle} follows a related but weaker allocation: its first RVQ layer receives temporal AR and later layers receive non-autoregressive refinement, without assuming that the residual order is a semantic hierarchy. These systems use long-context capacity for the variables they expose as long-range and cheaper conditional models where the remaining dependence is shorter or more ambiguous.

This view also prevents a category error. Replacing RVQ with one continuous latent does not remove the semantic-acoustic allocation problem. Replacing a flow decoder with discrete tokens does not make continuation coherent. Discreteness and continuity define the output interface; dependency horizon, conditional ambiguity, and streaming determine how that interface should be modeled.

Table~\ref{tab:worked-cases} applies the same questions to seven representative systems. This is a worked literature analysis, not experimental validation: the entries state which variables are explicitly exposed, where global context is allocated, and what remains for conditional generation.

\begin{table}[H]
\centering
\footnotesize
\caption{Worked comparison of representative representation-model pairs. ``Global signal'' means information explicitly available to the component with the longest context; it does not imply a universal semantic ontology.}
\label{tab:worked-cases}
\begin{tabularx}{\linewidth}{@{}lYYY@{}}
\toprule
System & Representation and global signal & Modeling allocation & Critical path \\
\midrule
AudioLM~\citep{audiolm} & w2v-BERT semantic tokens plus SoundStream RVQ & Semantic LM predicts a low-rate content trajectory first; acoustic LMs realize it conditionally. & temporal generation at successive stages \\\addlinespace[2pt]
VALL-E~\citep{valle} & EnCodec RVQ; text and prompt condition layer 1 & First codec layer gets temporal AR; residual layers get non-autoregressive refinement. & $N$ causal steps for RVQ-1, then $K-1$ layer-wise refinement passes \\\addlinespace[2pt]
MusicGen~\citep{musicgen} & EnCodec RVQ with text conditioning & Delayed AR prediction preserves a causal inter-codebook order without flattening all $NK$ tokens. & approximately $N+K-1$ positions \\\addlinespace[2pt]
UniAudio~\citep{uniaudio} & multi-codebook codec tokens with task conditioning & A global temporal transformer models frame history; a local transformer resolves within-frame codebook depth. & temporal steps plus local depth prediction \\\addlinespace[2pt]
SoundStorm~\citep{soundstorm} & semantic conditioning plus SoundStream tokens & Semantic conditioning plus bidirectional masked refinement parallelizes codec-token generation. & scheduled parallel refinement rounds \\\addlinespace[2pt]
DiTAR~\citep{ditar} & continuous acoustic frames conditioned by text and history & A causal patch-level LM sets context; a diffusion transformer resolves each patch's detail. & causal patch steps, each with diffusion evaluations \\\addlinespace[2pt]
CALM~\citep{calm} & continuous audio-VAE frames & A causal Transformer sets context; a consistency head replaces categorical prediction with continuous sampling. & causal frame steps, each with consistency-model evaluations \\
\bottomrule
\end{tabularx}
\end{table}

The table supports three limited conclusions. First, whether the interface is discrete or continuous does not determine the modeling scheme. Second, global information can be supplied by a latent stream, external conditioning, or causal history. Third, reducing critical-path evaluations changes which conditional dependencies a finite model must approximate. The table does not establish that one allocation is optimal; that requires the controlled comparisons proposed in Section~\ref{sec:open}.

\subsection{What should a unified audio model share?}

The preceding discussion uses speech because linguistic content makes long-range structure easy to name. A general audio model, however, may cover speech, music, environmental sound, and mixtures of them, as pursued by UniAudio~\citep{uniaudio} and Ming-UniAudio~\citep{minguniaudio}. Unification should not require these domains to share one fixed notion of semantics. It should provide a common interface for variables that play the same modeling role.

Let $g$ denote variables whose dependency horizon requires global modeling, and let $\ell$ denote the remaining realization detail: the same distinction as Principle~\ref{prin:entropy}'s $s_t$ and $a_t$, renamed here because ``semantic'' and ``acoustic'' are speech-specific labels that do not generalize cleanly across domains. In speech, $g$ may include linguistic content, dialogue state, speaker identity, and discourse-level prosody. In music, it may include melody, harmony, rhythm, instrumentation, and section structure. In environmental sound, it may describe event order, source persistence, and scene evolution. Singing, audio-visual generation, and other mixed settings can expose several of these variables at once. What these variables share is not an ontology. An error in $g$ changes the global continuation, whereas multiple values of $\ell$ may remain perceptually valid after $g$ and the conditioning are fixed.

\begin{table}[H]
\centering
\small
\caption{The same functional division appears across audio domains, but the variables occupying each role change with the domain and available conditioning.}
\begin{tabularx}{\linewidth}{@{}lYY@{}}
\toprule
Domain & Variables that may require long-context modeling & Conditionally ambiguous realization detail \\
\midrule
Speech & linguistic content, dialogue state, speaker, global prosody & excitation, phase, fricative texture, local room response \\
Music & melody, harmony, rhythm, form, instrumentation & micro-timing, performance variation, local timbral texture \\
Environmental sound & event sequence, source persistence, scene state & exact event waveform, background texture, reverberation \\
\bottomrule
\end{tabularx}
\end{table}

External conditioning can move the boundary. Text supplies much of the linguistic plan in TTS. Lyrics and symbolic controls can expose part of a musical plan. Video can constrain event timing and scene continuity. Without such conditioning, the audio model must generate those variables itself. The appropriate split between $g$ and $\ell$ is therefore conditional on both the domain and the information already provided.

We propose the following design hypothesis for unified audio systems. An architecture can share an encoder or tokenizer, temporal backbone, conditioning interface, and acoustic decoder without forcing every example into speech-style semantic tokens. Domain prompts, latent factors, specialized heads, or experts may expose different global variables and still retain a common generation mechanism. Variables that benefit from longer context should receive global modeling capacity. Detail that remains ambiguous under the available conditioning may be assigned to a local or iterative generator. Whether the same allocation works across speech, music, and environmental sound remains an empirical question.

\section{Open Problems in Audio Generative Modeling}
\label{sec:open}

The framework suggests a basic rule for future work: evaluate a representation and its downstream model as a pair. Existing codec surveys~\citep{survey} and benchmarks such as DASB~\citep{dasb} provide important reconstruction and downstream evidence, but the pairing question additionally requires a controlled generator and inference budget. A new latent should not be judged only by reconstruction, and a new generator should not be compared across unmatched latent rates or distortion levels. The useful question is whether the pair improves generation at matched perceptual quality, training compute, inference cost, and lookahead.

\subsection{Evaluate modelability under controlled conditions}

Modelability is currently inferred from downstream results obtained with different architectures and budgets. A controlled study should vary one side of the pair at a time. With the latent fixed, compare compatible temporal, depth, iterative, and hybrid schemes under matched compute and report downstream quality rather than raw losses across objectives. Within a representation family, hold the backbone fixed and vary frame rate, codebook depth, semantic supervision, or continuous dimension at matched reconstruction quality.

The experiment cannot observe the components of Equation~\ref{eq:factor-gap} separately, but it can measure their practical consequence under a fixed budget. Predictive or denoising loss should therefore be accompanied by semantic stability, speaker and prosody retention, perceptual quality, critical-path evaluations, and real-time cost. Reconstruction and generation must be reported separately. Short-form TTS is not a substitute for continuation, and offline throughput is not a substitute for streaming latency.

\subsection{Co-design the latent and its modeling strategy}

Most systems optimize a codec for reconstruction, freeze it, and train the generator afterward. This is convenient, but it leaves the downstream interface out of codec design. Co-design should decide whether capacity is placed in time, codebook depth, or feature dimension, and whether the downstream architecture provides the context and output distribution needed at that location. Co-design does not require end-to-end training. It first requires a shared objective and matched evaluation.

Naive end-to-end training is not sufficient. As the encoder changes, the generator sees a moving target. As predictability is rewarded, the encoder may discard difficult but perceptually important information. Useful approaches must constrain distortion and still improve the empirical profile in Equation~\ref{eq:modelability-profile}, for example through alternating updates, stop-gradient paths, or a slowly moving tokenizer. The decisive baseline is a frozen codec with the same generator compute and perceptual quality.

\subsection{Design the latent and model jointly for streaming}

Streaming is not an encoder property alone. It jointly constrains frame boundaries, latent state, modeling strategy, and sampling schedule. Audio's information density is non-stationary, so silence, sustained vowels, plosives, and music onsets need not receive the same rate. TADA~\citep{tada} and FlexiCodec~\citep{flexicodec} explore variable-rate representations, but removing the regular clock also changes positional encoding, buffering, and the downstream prediction problem.

A complete streaming design must specify the boundary policy, the information carried at each boundary, and the model that advances that state under bounded lookahead. For continuous latents, the corresponding experiment is a semantic-rate curve: compress the same causal SSL representation across dimensions and frame rates, then measure understanding, reconstruction, continuation, and real-time cost with matched downstream models. This would show whether a single continuous stream is sufficient or whether a small structured set of streams is needed.

\subsection{Decide what requires long-context modeling}

The semantic-acoustic split should not be fixed by convention. Different tasks have different long-range variables. Linguistic content dominates speech continuation. Melody, rhythm, instrumentation, speaker identity, room acoustics, or expressive prosody may instead remain global in other settings. The research question is which variables must be exposed before local acoustic generation, and where that boundary should appear: between tokenizers, between RVQ layers, between model stages (Section~\ref{sec:audiolm}), or between the outer and inner loops of a hybrid model (Section~\ref{sec:hybrids}).

Compatibility with text LLMs is one instance of this problem. Discrete tokens fit a softmax interface but can dominate context length and need not align with text semantics. Continuous features may align more naturally with hidden states but require projectors and a generative output head. A meaningful comparison should test understanding and generation in the same backbone under a fixed wall-clock budget. Interface convenience should not be mistaken for a well-placed semantic boundary.

\section{Conclusion}
\label{sec:conclusion}

The distinction between discrete and continuous latents describes the interface, but it is not sufficient to explain an audio generation system. The representation determines where rate, information, and uncertainty are placed. The modeling strategy determines which dependencies receive long-context prediction and how the remaining ambiguity is represented. These choices must be evaluated together under constraints on distortion, compute, and streaming.

The view that semantics is low-entropy and acoustics is high-entropy provides a useful starting intuition, but the more general diagnostic questions concern dependency horizon and conditional ambiguity. AudioLM~\citep{audiolm} exposes a semantic trajectory before acoustic realization. RVQ creates an ordered capacity axis without guaranteeing semantic order. How much of the remaining realization ambiguity a system can then represent depends on its modeling scheme, whether that is RVQ's chain rule, a Gaussian mixture, or a diffusion process, not on whether the latent is discrete or continuous. Continuous and hybrid systems place these same divisions at different locations. The four-objective design space and worked comparisons provide a common language for comparing these representation--model pairs and for deciding which part of the pair future work should change.

\small
\bibliographystyle{unsrtnat}
\bibliography{references}

\clearpage
\normalsize
\appendix
\begin{center}
{\LARGE\bfseries Appendix}
\end{center}
\vspace{0.75em}

\section{Why Audio Needs a Latent}
\label{app:background}

\subsection{Mel as the first widely used latent}

Generating audio in a latent space is older than learned codecs. The mel-spectrogram already acts as a handcrafted latent: it reduces temporal and frequency resolution, discards phase, preserves a perceptually useful spectral envelope, and delegates waveform realization to a vocoder. This was the standard TTS interface for years: a model predicted mel frames, and Griffin-Lim~\citep{griffinlim}, WaveNet~\citep{wavenet}, or HiFi-GAN~\citep{hifigan} converted them to waveform. Tacotron 2~\citep{tacotron2} is a canonical example of this pipeline. VITS~\citep{vits} moved toward a learned latent and integrated waveform decoder, but retained the same broad division between a lower-rate generative representation and waveform realization.

The mel scale itself predates any of this engineering by decades and was not designed for generative modeling at all. Stevens, Volkmann, and Newman~\citep{melscale} derived it from a 1937 psychoacoustic experiment: listeners adjusted a tone until it sounded subjectively half as high as a reference tone, and the resulting nonlinear mapping between perceived pitch and physical frequency defined the scale, named after \emph{melody}. Davis and Mermelstein~\citep{mfcc} turned this psychoacoustic curve into a practical speech feature, the mel-frequency cepstral coefficient, more than four decades before mel frames became a generation target. The representation that anchored the codec era therefore originated as a measurement of human pitch perception, not as an engineering compromise, and it carries that perceptual prior into every later system that adopts it.

\fig{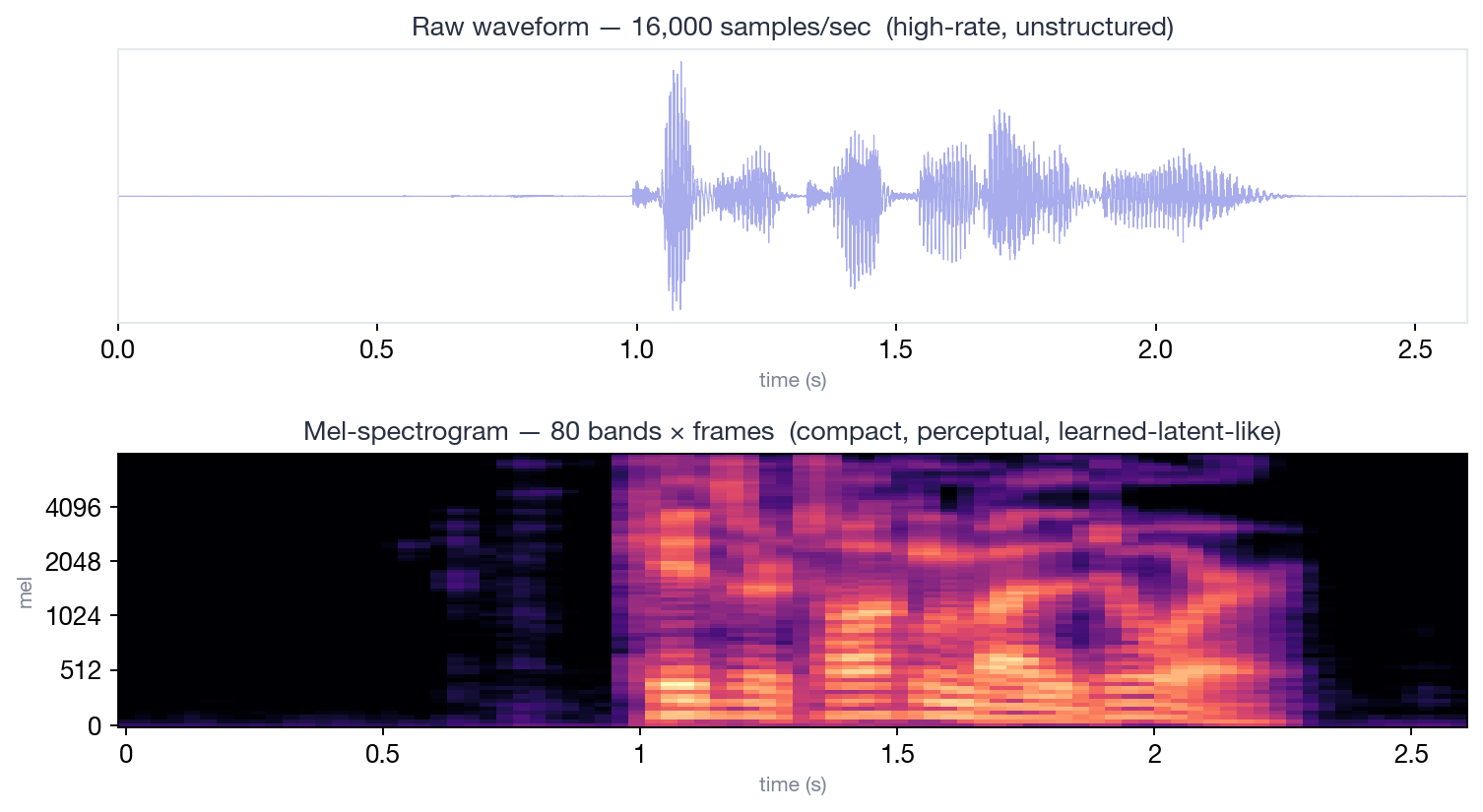}{0.95}{The mel-spectrogram is the original audio latent in practice: a lower-rate time--frequency scaffold that keeps the coarse structure of speech and leaves waveform realization to a vocoder.}

A typical mel representation turns a long waveform into a compact $N\times 80$ matrix. It retains spectral envelope, formants, and broad dynamics, and discards phase and fine waveform detail that a vocoder realizes later. Learned codecs and audio VAEs replace the handcrafted transform, but they inherit the same division: expose a lower-rate scaffold, then recover a plausible waveform.

Why not model the waveform directly? At 24~kHz, one second contains 24,000 samples. A 50~Hz latent contains only 50 frames. The difference determines context length, autoregressive steps, KV-cache cost, and whether listen-and-speak generation is practical. WaveNet~\citep{wavenet}, WaveGrad~\citep{wavegrad}, and DiffWave~\citep{diffwave} showed that neural waveform synthesis is possible, often as a conditional vocoder. Latent methods remain attractive because they keep waveform-rate generation out of the long-context model.

Audio also admits many perceptually acceptable realizations. Phase, micro-timing, room response, and noise texture can change without altering the relevant content or perceived naturalness. A pointwise waveform target therefore asks the generator to commit to details for which many answers are acceptable. Mel plus a vocoder handled this by discarding phase explicitly. Modern generative decoders make the division learnable: the latent carries a scaffold, and the decoder produces a perceptually valid realization.

Mel did not disappear once learned codecs and continuous latents became available; it kept reappearing inside them under a different role. dMel~\citep{dmel} quantizes mel bands directly rather than learning a new discrete bottleneck. Grad-TTS~\citep{gradtts} and DiffSinger~\citep{diffsinger} run diffusion directly over mel-like acoustic features rather than over a learned latent. Several neural codecs also keep an auxiliary mel-reconstruction loss alongside their adversarial and time-domain objectives. What kept mel useful was never that it was handcrafted; it was that it already had low dimensionality, strong frame-to-frame correlation, and content structure exposed at a low rate, which are precisely the properties a generative latent needs. Learned codecs replaced the handcrafted transform, but a number of the systems discussed in this paper still rely on some version of the same underlying scaffold.

Audio structure unfolds over several time scales. Speech meaning, musical form, speaker identity, and prosody can persist over seconds or minutes. A useful representation must therefore do more than compress local frames. It must decide which dependencies remain visible to the model responsible for long-range planning. This is why frame rate, semantic supervision, and streaming causality repeatedly reappear in the main text.

\subsection{What may be discarded, and what must remain}

The ambiguity removed by an audio latent is not uniform. Absolute phase and the precise realization of noise can often change without changing the perceived event. Harmonic relations, formant trajectories, pitch, and transient timing can instead be perceptually decisive. Even this distinction is conditional: phase coherence around a transient matters, although the absolute phase of an isolated steady component may not. A representation should therefore discard particular realizations of locally ambiguous detail without discarding the structure that constrains them.

Sound texture makes this distinction concrete. Rain, applause, fricative noise, and reverberation tails can admit many waveform realizations with similar time-averaged statistics~\citep{mcdermott}. Voiced harmonics and note onsets are less exchangeable. Unlike an image, however, audio does not place texture in one spatial region and structure in another. A fricative can follow a vowel within tens of milliseconds, and reverberation overlaps the next phoneme. The two may also occupy different frequency bands in the same frame. A fixed-rate latent must allocate capacity before it knows which kind of information a frame contains.

This observation sharpens the role of a generative decoder. Its purpose is not merely to invert compression. It should reconstruct constrained structure and sample a valid realization of detail that the latent intentionally leaves ambiguous. Mel plus a vocoder implements this division by hand. Adversarial, diffusion, and flow-based decoders learn versions of the same division. The quality of the interface depends on whether the discarded uncertainty is genuinely local and perceptually non-unique; if a decoder must invent linguistic content, melody, or speaker identity, the latent has hidden too much.

\section{From VQ-VAE to Modern Audio Codecs}
\label{app:codec}

VQ-VAE~\citep{vqvae} introduced the basic learned discrete bottleneck. SoundStream~\citep{soundstream} turned it into the modern neural audio codec template: a convolutional encoder and decoder, adversarial and spectral reconstruction losses, and residual vector quantization. EnCodec~\citep{encodec} made the template broadly reusable and streamable. DAC~\citep{dac} increased model and discriminator capacity and pushed perceptual reconstruction further. Together these systems established the interface later consumed by codec language models.

RVQ increases capacity in depth. With frame rate $r$, $K$ codebooks, and codebook sizes $|\mathcal Z_k|$, its nominal bitrate is
\begin{equation}
R_{\mathrm{disc}}=r\sum_{k=1}^{K}\log_2|\mathcal Z_k|.
\end{equation}
Table~\ref{tab:codec-specs} reports this quantity for representative codecs, which place the same nominal bitrate at different points along the frame-rate-versus-codebook-depth trade-off. EnCodec~\citep{encodec} at 75~Hz with two active 1024-entry codebooks gives exactly 6~kbps. Flattening the stream produces 150 token decisions/s. Predicting the two codebooks in parallel instead leaves 75 temporal steps/s but creates a two-variable conditional problem within each frame. The same equation explains why Mimi~\citep{moshi} reaches 1.1~kbps at 12.5~Hz with eight codebooks, why DAC~\citep{dac} instead uses 86~Hz and nine codebooks, and why WavTokenizer~\citep{wavtokenizer} and BigCodec~\citep{bigcodec} push capacity entirely into a single, larger codebook rather than into depth. Raising the frame rate and adding codebook depth are two ways of buying the same nominal bitrate, but they are not interchangeable for a downstream model: one shortens the temporal sequence at the cost of a deeper per-frame problem, and the other keeps a flat one-token-per-step interface at a higher temporal rate.

\begin{table}[H]
\centering
\small
\caption{Representative discrete audio codecs place the same nominal bitrate at different points along the frame-rate-versus-codebook-depth trade-off.}
\label{tab:codec-specs}
\begin{tabularx}{\linewidth}{@{}lYYY@{}}
\toprule
Codec & Frame rate & Codebook configuration & Nominal bitrate \\
\midrule
SoundStream~\citep{soundstream} & 75~Hz & up to 24 codebooks (first two 1024-entry, later 128-entry) & 3--18~kbps, scalable \\
EnCodec~\citep{encodec}, 24~kHz & 75~Hz & up to 32 $\times$ 1024-entry codebooks & 1.5--24~kbps, scalable \\
DAC~\citep{dac}, 44.1~kHz & 86~Hz & 9 $\times$ 1024-entry codebooks & $\approx$8~kbps \\
Mimi~\citep{moshi} & 12.5~Hz & 8 $\times$ 2048-entry codebooks & 1.1~kbps \\
WavTokenizer~\citep{wavtokenizer} & 75~/~40~Hz & 1 $\times$ 4096-entry codebook & 0.9~/~0.48~kbps \\
BigCodec~\citep{bigcodec} & 80~Hz & 1 $\times$ 8192-entry codebook & 1.04~kbps \\
\bottomrule
\end{tabularx}
\end{table}

RVQ guarantees residual order, not semantic order. Earlier codebooks reduce the largest remaining reconstruction error. They are not forced to represent phonemes, prosody, or any other named factor. SpeechTokenizer~\citep{speechtokenizer} and Mimi~\citep{moshi} add distillation precisely because reconstruction training alone does not guarantee that the first layer is linguistically useful. FACodec~\citep{ns3} takes the more explicit route of assigning content, prosody, timbre, and acoustic detail to separately supervised factors.

Alternatives change where the cost is paid. WavTokenizer~\citep{wavtokenizer} and BigCodec~\citep{bigcodec} remove depth unrolling but use relatively high token rates and large per-token capacity. SNAC~\citep{snac} distributes streams across temporal resolutions. FSQ~\citep{fsq} quantizes scalar dimensions to fixed levels and avoids a learned embedding codebook. Comparisons should include reconstruction, frame rate, token decisions, dependency-horizon accessibility, and the downstream architecture.

\subsection{A reconstructable token stream may still be difficult to model}

Codec training optimizes an encoder-decoder pair. The encoder need only transmit information that its paired decoder cannot recover from architectural priors and local context. Convolutional locality, periodic inductive biases, and a powerful waveform generator may therefore permit a compact but highly entangled code. Such a code can be an efficient interface to that decoder without being a convenient target for a transformer trained later.

This is the precise sense in which a reconstruction-only token stream can become a private language of its decoder. The phrase does not mean that the tokens contain no linguistic information. It means that information useful to an external model may be distributed across time, codebook depth, and decoder-specific correlations. A linear probe or finite transformer may fail to access it even when an unconstrained decoder reconstructs intelligible speech. Reconstruction distortion therefore upper-bounds neither dependency-horizon accessibility nor downstream generation quality.

Several tokenizer families can be read as attempts to make the interface more public. SpeechTokenizer~\citep{speechtokenizer} and Mimi~\citep{moshi} distill SSL features into selected codec layers. FACodec~\citep{ns3} assigns supervision to content, prosody, and timbre factors. MOSS-Audio-Tokenizer~\citep{mossaudiotokenizer} adds language-model-based understanding tasks so that token structure is useful beyond reconstruction. These objectives differ, but all constrain which information must be directly accessible before the waveform decoder is applied.

The appropriate test is consequently paired. First measure reconstruction with ground-truth latents. Then freeze the representation and train a specified downstream model under a fixed budget. Evaluate likelihood or denoising loss, semantic probes, conditional generation, and free-running stability. A representation that wins only the first test has demonstrated codec quality, not modelability.

\section{How Audio Models Consume Codec Tokens}
\label{app:downstream}

A multi-codebook codec does not define a unique token sequence. A sequential model must choose an order over time $t$ and codebook depth $k$. A parallel or iterative model must instead choose which variables to group. A fully flattened order is simple but multiplies sequence length by $K$. Predicting all codebooks at one time step is shorter, but assumes conditional independence or requires a second model within the frame.

VALL-E~\citep{valle} places long-range autoregression on the first codec layer and predicts the remaining layers non-autoregressively. MusicGen~\citep{musicgen} uses a delay pattern so several codebooks can be predicted in parallel and still preserve a causal dependency order. UniAudio~\citep{uniaudio} and Moshi~\citep{moshi} separate the two axes more explicitly: a temporal transformer models frames, and a smaller depth transformer models codebooks inside each frame. These are not merely implementation tricks. Each one states where the system believes the hard conditional dependencies live.

Discrete tokens do not require autoregressive generation. SoundStorm~\citep{soundstorm} and MAGNeT~\citep{magnet} use masked, iterative generation over codec tokens. Conversely, autoregression does not require discrete tokens: a continuous frame can be generated from a conditional density, diffusion, flow, or consistency head. The important choices are the context allocation and conditional distribution model, not whether the latent happens to be categorical.

\subsection{Two axes, several valid factorizations}

For RVQ tokens $z_t^{1:K}$, the exact chain rule can be written as
\begin{equation}
p(z)=\prod_{t=1}^{N}\prod_{k=1}^{K}
p\!\left(z_t^k \mid z_{<t}^{1:K},z_t^{<k}\right).
\end{equation}
This equation specifies dependencies but not one mandatory architecture. A flattened LM implements the order directly and pays roughly $NK$ serial token decisions. A temporal model with a local depth transformer preserves the same broad dependency pattern with $N$ outer steps and a smaller within-frame process. A delay pattern changes which codebooks can be emitted together. A masked model replaces a causal order with an iterative conditional approximation.

\begin{table}[H]
\centering
\small
\caption{Common ways to consume a $K$-codebook latent of length $N$. Critical-path evaluations are schematic because implementations can cache, parallelize, or group predictions differently.}
\begin{tabularx}{\linewidth}{@{}lYYY@{}}
\toprule
Scheme & Dependency retained & Approximate critical path & Representative systems \\
\midrule
Flatten time and depth & Full chosen chain rule & $NK$ & generic codec LM \\
First layer AR, later layers NAR & Long context mainly in layer 1 & $N$ plus refinement & VALL-E~\citep{valle} \\
Delayed parallel streams & Shifted cross-codebook context & about $N+K$ & MusicGen~\citep{musicgen} \\
Temporal plus local depth model & Frame history and within-frame order & $N$ outer steps & UniAudio~\citep{uniaudio}, Moshi~\citep{moshi} \\
Masked iterative prediction & Bidirectional visible context & refinement rounds & SoundStorm~\citep{soundstorm}, MAGNeT~\citep{magnet} \\
\bottomrule
\end{tabularx}
\end{table}

The comparison exposes two different costs. Critical-path evaluations measure how many dependent network calls are required. Approximation difficulty describes what conditional structure the chosen scheme asks a finite model to learn or ignore. Parallelizing codebooks reduces the first cost but can enlarge the second. Flattening does the reverse. Neither quantity predicts latency without model size and hardware. This is why codebook depth cannot be evaluated independently of the unrolling scheme.

\section{The Fixed-Rate Grid and Content-Adaptive Latents}
\label{app:variable-rate}

Most audio latents emit a fixed number of frames per second. This makes batching, positional encoding, and causal synchronization simple, but it allocates equal temporal capacity to silence, steady vowels, transients, and rapid phonetic changes. Their local information rates are not equal. RVQ quantizer dropout changes bits per frame, and multi-scale codecs change resolution across streams, but neither necessarily changes the number of temporal decisions seen by the downstream model.

A content-adaptive latent instead asks where a new representation element is needed. Vision tokenizers such as TiTok~\citep{titok} and FlexTok~\citep{flextok} show that a generative representation need not preserve a dense fixed grid. Variable-rate representation learning~\citep{varrate} treats audio token count as a learned resource. ALMTokenizer~\citep{almtokenizer} uses query-based compression for a low-rate semantic-rich codec. TADA~\citep{tada} aligns acoustic units with text units, replacing a uniform acoustic clock with a text-conditioned clock. FlexiCodec~\citep{flexicodec} merges semantically similar neighboring frames and permits the operating frame rate to vary. These approaches differ in whether they require text, whether boundaries are causal, and whether the downstream model consumes the variable-rate sequence directly.

The last distinction is crucial. A codec can use adaptive segmentation internally and then expand its output back to a regular grid before decoding. That may improve compression without shortening the sequence modeled by an LM. For audio generation, the stronger goal is a latent whose external sequence length changes with content and whose timing remains recoverable. Such a representation also needs positional information that distinguishes elapsed time from token index.

Variable rate and streaming are not logical opposites, but they compete for information. A causal segmenter must decide whether to close a segment without seeing its future, so stable boundaries may require buffering. Text-aligned systems obtain a clock from the conditioning sequence but do not directly generalize to untranscribed music or environmental audio. Audio-only systems can use change detection or learned boundary probabilities, but must expose duration and maintain synchronization. The unresolved problem is therefore not merely adaptive compression. It is a causal, content-adaptive representation that a downstream generator can consume without first restoring a fixed grid.

\end{document}